\documentclass[lettersize,journal]{IEEEtran}
\usepackage{amsmath,amsfonts}
\usepackage{algorithmic}
\usepackage{algorithm}
\usepackage{array}
\usepackage[caption=false,font=normalsize,labelfont=sf,textfont=sf]{subfig}
\usepackage{textcomp}
\usepackage{stfloats}
\usepackage{url}
\usepackage{verbatim}
\usepackage{graphicx}
\usepackage{cite}
\usepackage{tikz}
\usepackage{xcolor}
\usepackage{datetime}
\usepackage{lipsum}
\usepackage{mathtools}
\usepackage{cuted}
\usetikzlibrary{positioning}
\usepackage{balance}
\newcommand{\nint}[1]{\ensuremath\left\lfloor#1\right\rceil}

\newcommand\copyrighttext{%
	\footnotesize \textcopyright This work has been submitted to the IEEE for possible publication. Copyright may be transferred without notice,after which this version may no longer be accessible.}
\newcommand\copyrightnotice{%
	\begin{tikzpicture}[remember picture,overlay]
		\node[anchor=north,yshift=-10pt] at (current page.north) {\fbox{\parbox{\dimexpr\textwidth-\fboxsep-\fboxrule\relax}{\copyrighttext}}};
	\end{tikzpicture}%
}

\begin{document}

\title{Golden Modulation: a New and Effective Waveform for Massive IoT}

\author{Lorenzo Vangelista,~\IEEEmembership{Senior Member,~IEEE}, Bruno Jechoux, Jean-Xavier Canonici and Michele Zorzi,~\IEEEmembership{Fellow,~IEEE}
\thanks{A short (4-page) version of this manuscript has been submitted to IEEE MedComNet 2023.}}



\maketitle


\copyrightnotice

\begin{abstract}
This paper considers massive Internet of Things systems, especially for LoW Power Wide Area Networks, that aim at connecting billions of low-cost devices with multi-year battery life requirements. Current systems for massive Internet of Things exhibit severe problems when trying to pursue the target of serving a very large number of users.
In this paper, a novel asynchronous spread spectrum modulation, called Golden Modulation, is introduced. This modulation provides a vast family of equivalent waveforms with very low cross-interference even in asynchronous conditions, hence enabling natural multiuser operation without the need for inter-user
synchronization or for interference cancellation receivers. Thanks to minimal interference between
waveforms, coupled with the absence of coordination requirements, this modulation can accommodate very high system capacity. The basic modulation principles, relying on spectrum spreading via
direct Zadoff-Chu sequences modulation, are presented and the corresponding theoretical
bit error rate performance in an additive white Gaussian noise channel is derived and compared by simulation with realistic
Golden Modulation receiver performance. The demodulation of the Golden Modulation is also described, and its performance in the presence of uncoordinated multiple users is characterized.  \end{abstract}


\parskip 2ex

\section{Introduction}

\IEEEPARstart{M}{a}ssive access for the Internet of Things (IoT) has been investigated for quite some time (see, e.g., \cite{zucchetto}) and is a hot research topic for current and future systems, such as 6G (see, e.g., \cite{gao2022massive}). Solutions avoiding the need for device coordination (which typically requires significant energy and communications resources) are considered as particularly promising (see, e.g., \cite{dobre}). 
For IoT systems operating in the licensed spectrum, of which NB-IoT \cite{nb-iot} is currently the most popular solution, the research focuses on access schemes, and the waveform design is based on variations of multi-carrier modulations (see \cite{banelli} in \cite{maric2022information}). 
Instead, the waveforms used for IoT systems operating in the unlicensed spectrum exhibit a wide variety. Traditional solutions range from the usual Cyclic Prefix-OFDM multi-carrier modulation  for the DECT-2020 New Radio \cite{dect} to ultranarrowband  D-BPSK (Differential Binary Phase-Shift Keying) for the SigFox systems \cite{sigfox}. 
A new spread spectrum waveform emerged with the LoRa/LoRaWAN\textsuperscript{\textregistered} system \cite{vangelista-2017, Chiani2019}, which gained strong momentum and, at the time of writing, is worldwide the most popular system for massive IoT in unlicensed bands. The wide deployment of LoRaWAN networks highlighted some critical aspects related specifically to situations in which the density of End Nodes (ED) is very high \cite{radio-access-lora} \cite{alouini}.  The ability of LoRa/LoRaWAN to multiplex concurrent data transmissions from the EDs relies on the availability of different ``orthogonal'' waveforms, also referred to as logical channels. When the density of EDs becomes high, the number of logical channels becomes a bottleneck, making the network congested. The fact that the ``orthogonal'' waveforms are actually not fully orthogonal (see \cite{Benkhelifa-2022}) makes the problem even worse. 

From the above discussion, it emerges that a major open problem in massive access for the Internet of Things (IoT) lies fundamentally in the design of the waveform, which must be such that a very large number of distinguishable (ideally orthogonal) channels are available. In the LoRa systems, in each frequency channel separate logical channels are obtained by using different Spreading Factors (defined as the number of transmitted samples per information symbol \cite{vangelista-2017, Chiani2019}), whose number is limited to 6 since they can take a value in the set $\left\{7, 8, \ldots, 12 \right\}$ \cite{lorawan_l2}. It must be pointed out that the research for good linear frequency modulated signals, such as those used in LoRa, is still active \cite{kadambi}.

The main contribution of this paper is the design and evaluation of a novel set of waveforms, called Golden Modulation (GM), which departs conceptually from linear frequency modulated signals, a characteristic of the LoRa Modulation, and allows a huge increase in the number of parallel channels, compared to the state of the art mentioned above.  GM provides both leading edge sensitivity and very high multi-user capacity. 
This modulation supports flexible bandwidth and spreading factor 
as well as very good uncoded sensitivity. In addition, it provides a vast family of equivalent  waveforms with very low interference between them even in asynchronous conditions. More precisely, it achieves very low interference between non-coordinated users, hence enabling natural multiuser operation without the need for inter-user synchronization or for interference cancellation receivers. 
Thanks to minimal interference between waveforms, coupled with the absence of coordination requirements, this modulation can accommodate very high system capacity simply relying on the  multiplexing between autonomously operating users, separated by their waveforms, freely transmitting whenever they need to. 
As a comparison, LoRa modulation provides one waveform per spreading factor while hundreds or even thousands of available waveforms  (see Table \ref{tab:table1}) are possible using Golden Modulation. 

GM is based on a novel and generalized perspective on the use of the well known Zadoff-Chu sequences. This feature is enabled by using the Zadoff-Chu sequences to carry information rather than (as it is usually done) for synchronization and/or channel estimation in the headers of a packet in a digital communication link (see, e.g., \cite{kaushik}). In particular, although the use of Zadoff-Chu sequences was recently discussed for signaling in IoT scenarios \cite{ZCNET-2022,Deparis-2020}, we are the first to recognize their potential to provide a large number of distinct logical channels, thereby providing an effective solution for true IoT massive access. The efficient implementation of GM is also described in this paper at both the transmitter and the receiver sides. 

The paper is organized as follows. In Section \ref{sec:iotsystem} the system requirements for Massive IoT systems are reviewed, with emphasis on the effect of network synchronization on the possibility to fulfill the requirements. In Section~\ref{sec:notation}, the notation used in the paper is introduced. 
In Section \ref{sec:sequences}, a short review and description are provided for the  frequency shifted Zadoff-Chu Sequences, which are employed in GM. In Section~\ref{sec:single-link}, the use of GM in a single user system is presented.  Section \ref{sec:receiver} describes the GM receiver.
In Section \ref{sec:multiple-links}, the use of GM in the general setting of multiple users with multiple spreading factors is presented. In Section~\ref{sec:higher_layers_networks}, GM is framed in the general perspective of a system including higher layers on top of the physical layer. Finally, conclusions are drawn in Section~\ref{sec:conclusions}.

\section{Massive IoT System Requirements}
\label{sec:iotsystem}
Massive IoT aims at connecting billions of low-cost devices with multi-year battery life. To achieve such stringent requirements, some key bottlenecks must be removed: in particular, the system capacity shortage faced in current LPWAN networks, both in unlicensed and licensed spectrum, must be solved. An overall capacity several orders of magnitude higher than what is available in current state-of-the-art LPWANs is required in order to enable the expected number of connections per IoT gateway.
In addition, the device's power consumption must be minimized to enable multi-year battery life. The device power consumption is driven on one side by its required transmit power and on the other side by the energy required for synchronization onto the network. 
The required degree of network synchronization varies drastically depending on the physical layer and associated link layer of the system: OFDM-based systems such as NB-IoT, LTE cat M or the recently introduced DECT 2020 NR require stringent time and frequency synchronization among all devices, coupled with strict resource request and assignment mechanisms in order to maintain orthogonality of the PHY signals and to avoid packet collisions. This has a significant impact not only in terms of energy consumption but also in terms of protocol overhead, since specific channels for downlink synchronization, resource request and resource grant are then required, and consume radio resources. On the contrary, non-cellular LPWAN systems such as LoRa or SigFox can operate nearly without network synchronization. 
It is interesting to note that, due to the characteristics of massive IoT traffic, where each device transmits infrequently small amounts of data, the energy needed for network synchronization may become dominant over the energy required to effectively transmit data packets. This is especially true for cellular-based systems such as NB-IoT (also known as 5G IoT) or LTE cat M, where the device is a slave to the network which fully controls the access to the medium, implying high power consumption and high control overhead, hence reducing further the effective capacity of the system. A comparison between LoRaWAN and NB-IoT has been performed in \cite{buratti}, showing that LoRaWAN and NB-IoT have similar performance in terms of network capacity, with NB-IoT slightly outperforming LoRaWAN for small cells, at the cost of a significantly higher energy consumption even without considering the energy cost of network synchronization. The impact of NB-IoT synchronization on the energy budget for each NB-IoT Uplink data packet transmission in real conditions has been analyzed in \cite{Yang}. 
Overall, it seems reasonable to state that a scalable massive IoT system must avoid device network synchronization to make the device operation as autonomous as possible. It is hence of crucial importance to have a PHY layer able to operate with minimum or no device synchronization requirements, while ensuring in addition high sensitivity and high capacity.

\section{Notation}
\label{sec:notation}
For the sake of readability, in this section we provide a summary of the notation used in this paper.

A generic signal in the continuous time domain is denoted as $x(t), t\in \mathbb{R}$, where $\mathbb{R}$ is the set of real numbers and $x(t) \in \mathbb{R}$ or $x(t) \in \mathbb{C}$, where $\mathbb{C}$ is the set of complex numbers.

A generic discrete time signal $x_d(n)$ can be denoted as $x(nT_c)$ where $n \in \mathbb{Z}$ and $T_c$ is the temporal interval separating each sample of the signal. While we are well aware that discrete time signals are often denoted (with a slight abuse of notation) simply as $x(n)$, our notation is useful in this paper since we will be dealing with discrete signals with different temporal intervals. Dropping $T_c$ would possibly create confusion, while adding it improves readability at a minimal cost in terms of complexity of the notation.

Our notation is useful also when dealing with signals characterized by ``temporal blocks." We refer to the situation where a signal is made by appending sets of samples of cardinality $N$. For example, each frame can result from the modulation of a signal $s(\ell T)$ carrying a symbol from the set $\mathcal{S}$ (e.g., QAM symbols) in the $\ell$\textsuperscript{th} symbol time $T$: the temporal quantum between different signal samples of the modulated signal is $T_c=\frac{T}{N}$ where $N$ is the interpolating factor. So we can write the modulated signal as $x(nT_c)$.
Our notation is particularly useful when we want to refer to a certain part of the signal,
such as the part of the signal modulated by the $\ell$\textsuperscript{th} symbol: we can denote it as $x(\ell T+kT_c)$, $k \in \left\{0,1, \ldots, N-1 \right\}$, instead of the more usual notation $x_d(\ell N+k)$ $k \in \left\{0,1, \ldots, N-1 \right\}$. On the other hand, when no temporal information is involved,
we use the common notation where a sequence (such as a Zadoff-Chu sequence) is simply denoted by $x_d(n)$.

We refer the reader to~\cite{Cariolaro2011} for a general treatment of  signal theory according to our notation, and to~\cite{vangelista2022} for a paper on LoRaWAN using the proposed notation. Table~\ref{tab:notation} provides a reference summary of our notation, for the convenience of the reader.

Finally, in the rest of the paper, we denote as \textit{link} the communication between a single transmitter and a (single) receiver.
\begin{table}[!t]
\renewcommand{\arraystretch}{1.3}
\caption{A summary of the notation}
\label{tab:notation}
\centering
\begin{tabular}{c||c}
\hline
\bfseries symbols & \bfseries explanation\\
\hline\hline
$T_c$ & basic temporal interval between \\ & two samples of a digital signal\\
\hline
$T=N\cdot T_c$ & "block" interval of $N$ basic temporal intervals\\
\hline
$x(\ell T + k T_c)$ & signal at the index $\ell \cdot N + k$ \\
\hline
$x_d( k )$ & sequence at the index $k$ \\
\hline
\end{tabular}
\end{table}
\section{The  frequency shifted Zadoff-Chu Sequences}
\label{sec:sequences}
According to \cite{Chu-1972}, we can express the frequency shifted Zadoff-Chu Sequences
as follows

\begin{equation}
	Z_N^{r,q}(k) = 
	\begin{cases}
		&  \!\!  \!\!  e^{\jmath \frac{\pi}{N} r \left(k+1+2q\right)k} \; \text{for $N$ odd} \\
	    &  \!\!  \!\! e^{\jmath \frac{\pi}{N} r \left(k+2q\right)k} \; \text{for $N$ even}
	\end{cases}
\label{eq:defZC}
\end{equation}
where $q \in \mathbb{Z}$ , $N \in \mathbb{N}, N \neq 0$, $r \in \mathbb{N}$, $\gcd(r,N)=1$, and $k=0, \ldots, N-1$.
Usually, $r$ is called the ``root'' of the sequence and $N$ its length. It must be noted that, although in general  $q \in \mathbb{Z}$, meaningful values for $q$ are in the set $\left\{0,1, \ldots ,N-1\right\}$. For example, when $q \ge N$, Equation (\ref{eq:defZC}) becomes
\begin{equation}
	Z_N^{r,q}(k)  \!\! = \!\!
	\begin{cases}
		&   \!\!  \!\! \!\! \!e^{\jmath \frac{\pi}{N} r \left(k+1+2q\right)} \!=  e^{\jmath \frac{\pi}{N} r \left(k+1+2(q \!\! \!\! \! \mod \!\!  N)\right)k}  \; \text{for} \, N \, \text{odd}\\
		&  \!\!  \!\! \!\!  e^{\jmath \frac{\pi}{N} r \left(k+2q\right)} \!= e^{\jmath \frac{\pi}{N} r \left(k+2(q \!\! \!\! \! \mod \!\!  N)\right)k}  \; \text{for} \, N \, \text{even}
	\end{cases}
	\label{eq:defZC_mod}
\end{equation}
where clearly $(q  \!\!  \mod \!  N) \in \left\{0,1, \ldots , N-1\right\}$.

From Equations (9) and (10) in \cite{Chu-1972}, we have that the cyclic autocorrelation of the sequence $Z_N^{r,q}(\cdot)$  is non-zero only for a shift equal to 0.
From \cite{Popovic-1992}, we have that by selecting $N$ odd and any pair of roots $r_1, r_2$ such that $\gcd(r_1-r_2,N)=1$, the cyclic cross-correlation between the two Zadoff-Chu sequences $Z_N^{r_1,q}(\cdot)$ and $Z_N^{r_2,q}(\cdot)$ is equal to $\frac{1}{\sqrt{N}}$.  
In particular, if $N$ is a prime number, $(r_1-r_2)$ is always relatively prime to $N$, whichever the selected pair of roots $(r_1$,$r_2)$,\footnote{We recall that $r_1 <N$ and $r_2<N$ and that if $(r_1-r_2)<0$, in this context $(r_1-r_2)$ is equivalent to $(N+(r_1-r_2))$.} and the whole set of possible Zadoff-Chu roots can be used with minimal cross-correlations between the corresponding sequences.

\section{The Golden Modulation: single link description}
\label{sec:single-link}

\subsection{Discrete Time Signal Analysis }
\label{subsec:discrete-time-signal}

We suppose to have a source of $M$-ary symbols $M < N$ (see Section \ref{sec:sequences} and Equation (\ref{eq:defZC_mod})), emitting symbols at rate $R$. We indicate the generic symbol in the symbol period $\ell T$ as $q(\ell T)=q_\ell\in\left\{0,1, \ldots , M-1\right\}$, $T=1/R$, $\ell \in \mathbb{N}$. We denote  the ``chip'' period as $T_c=T/N$.  Then, the discrete-time modulated GM signal in the symbol period $\ell T$ is 
\begin{equation}
	x(\ell T + kT_c) \stackrel{\text{def}}{=} Z_N^{r,q_\ell}(\ell N + k) \; , \; k=0,\ldots,N-1
	\label{eq:1}
\end{equation}
In order to obtain an expression of the modulated signal highlighting its dependence on the set $\left\{q_\ell, N, r \right\}$ defining the parameters and the symbol, we rewrite Eq. (\ref{eq:1}), providing an alternative expression for the modulated signal,  as  
\begin{equation}
	x(\ell T + kT_c;q_{\ell},N;r) \stackrel{\text{def}}{=} Z_N^{r,q_\ell}(\ell N + k) \; .
	\label{eq:2}
\end{equation}
For a generic time $nT_c$, defining $\ell(n) \stackrel{\text{def}}{=}\lfloor \frac{n}{N}\rfloor$ and $k(n) \stackrel{\text{def}}{=} n \mod N$, we can write the discrete time modulated signal as 
\begin{equation}
	x(nT_c;q_\ell,N;r) = x_d(\ell(n) T + k(n)T_c;q_{\ell(n)},N;r).
\end{equation}
These alternative expressions for the modulated signals will be needed in the following, depending on the context, for example to assess the orthogonality or in Section~\ref{sec:multiple-links}.

As in \cite{Chiani2019}, we set $T_c=\frac{1}{B}$, where $B$ is the minimum bandwidth occupied by a GM link. As a consequence, the symbol duration $T$ is $T=\frac{N}{B}$, so that the symbol rate is $R=\frac{B}{N}$ and the bit rate $R_b= b \cdot  \frac{B}{N}, b=\log_2 M$. We should note that by selecting $N$ as a power of 2, we can define by analogy the Spreading Factor for GM as
\begin{equation}
	SF \stackrel{\text{def}}{=} \log_2 N \;.
	\label{eq:SF_GM}
\end{equation} 
If $N$ is not a power of 2, $SF$ in Equation~(\ref{eq:SF_GM}) results to be a non integer number; in that case we define 
\begin{equation}
	SF \stackrel{\text{def}}{=} \nint{\log_2 N} \;,
	\label{eq:SF_GM_no prime}
\end{equation} 
where the operator $\nint{\cdot}$ performs the rounding to the closest integer. 

GM demodulation is based on the orthogonality of Zadoff-Chu sequences with the same root and offsets $q,q'$. We have that (see Section \ref{sec:sequences})
\begin{equation}
\begin{split}
& \sum_{k=0}^{N-1} x(\ell T + kT_c;q,N;r) \cdot x^*(\ell T + kT_c;q',N;r)  = \\
& = \begin{cases}{N} & q=q' \\ 0 & q \neq q'\end{cases}
\end{split}
\label{eq:orthogonality}
\end{equation}
so the discrete-time signal space for the GM modulation admits as an orthonormal basis the set of sequences $\frac{1}{\sqrt{N}}x(kT_c;q,N;r)$ indexed by the possible transmitted symbol $q \in \{0,1, \ldots, M-1\}$ and for the running time index $kT_c$, $k \in \{0,1, \ldots, N-1\}$. We then observe that we are in the same conditions as in \cite{Robert2018}; by re-using the same arguments and the proof as  in \cite{Robert2018}, Sections II and III-A, in our case (see the Appendix for details), the BER performance in an Additive White Gaussian Channel (AWGN) can be accurately approximated by 
\begin{equation}
\ P_{b} \approx 0.5 \,Q\left(\sqrt{\Gamma \, 2^{SF+1}}-\sqrt{1.386\,SF +1.154}\right)
\label{eq:ber_lora}
\end{equation}
where $\Gamma$ is the Signal-to-Noise Ratio (SNR) and $Q(x)=\int_x^{\infty} \frac{e^{-a^2/2}da}{\sqrt{2\pi}}$ is the complementary Gaussian distribution function.

\subsection{Continuous Time Signal Analysis }
\label{subsec:continuous-time-signal}
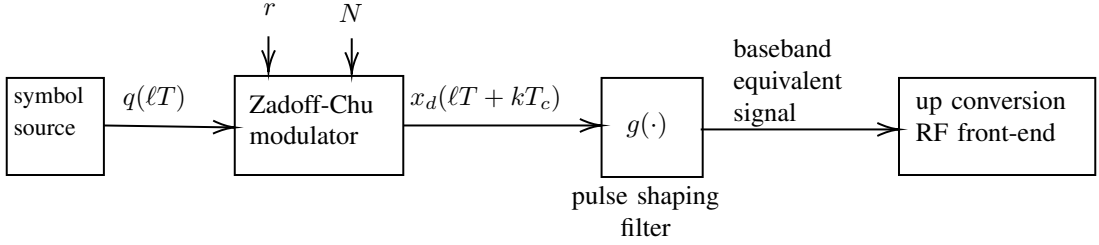
\begin{figure*}[htb]
	\centering

\tikzset{every picture/.style={line width=0.75pt}} 

\begin{tikzpicture}[x=0.75pt,y=0.75pt,yscale=-1,xscale=1]
	
	\draw   (101,101) -- (150,101) -- (150,150) -- (101,150) -- cycle ;
	\draw    (150,125) -- (214,125.97) ;
	\draw [shift={(216,126)}, rotate = 180.87] [color={rgb, 255:red, 0; green, 0; blue, 0 }  ][line width=0.75]    (10.93,-3.29) .. controls (6.95,-1.4) and (3.31,-0.3) .. (0,0) .. controls (3.31,0.3) and (6.95,1.4) .. (10.93,3.29)   ;
	\draw   (216,100) -- (301,100) -- (301,150) -- (216,150) -- cycle ;
	\draw    (233,80) -- (233,96) ;
	\draw [shift={(233,98)}, rotate = 270] [color={rgb, 255:red, 0; green, 0; blue, 0 }  ][line width=0.75]    (10.93,-3.29) .. controls (6.95,-1.4) and (3.31,-0.3) .. (0,0) .. controls (3.31,0.3) and (6.95,1.4) .. (10.93,3.29)   ;
	\draw    (275,81) -- (275,97) ;
	\draw [shift={(275,99)}, rotate = 270] [color={rgb, 255:red, 0; green, 0; blue, 0 }  ][line width=0.75]    (10.93,-3.29) .. controls (6.95,-1.4) and (3.31,-0.3) .. (0,0) .. controls (3.31,0.3) and (6.95,1.4) .. (10.93,3.29)   ;
	\draw    (301,125) -- (399,125) ;
	\draw [shift={(401,125)}, rotate = 180] [color={rgb, 255:red, 0; green, 0; blue, 0 }  ][line width=0.75]    (10.93,-3.29) .. controls (6.95,-1.4) and (3.31,-0.3) .. (0,0) .. controls (3.31,0.3) and (6.95,1.4) .. (10.93,3.29)   ;
	\draw   (401,101) -- (452,101) -- (452,151) -- (401,151) -- cycle ;
	\draw    (451,127) -- (549,127) ;
	\draw [shift={(551,127)}, rotate = 180] [color={rgb, 255:red, 0; green, 0; blue, 0 }  ][line width=0.75]    (10.93,-3.29) .. controls (6.95,-1.4) and (3.31,-0.3) .. (0,0) .. controls (3.31,0.3) and (6.95,1.4) .. (10.93,3.29)   ;
	\draw   (551,101) -- (650,101) -- (650,150) -- (551,150) -- cycle ;
	
	\draw (103,104) node [anchor=north west][inner sep=0.75pt]   [align=left] {{\fontfamily{ptm}\selectfont {\small symbol }}\\{\fontfamily{ptm}\selectfont {\small source}}};
	\draw (158,103) node [anchor=north west][inner sep=0.75pt]   [align=left] {$\displaystyle q( \ell T)$};
	\draw (222,108) node [anchor=north west][inner sep=0.75pt]   [align=left] {{\fontfamily{ptm}\selectfont Zadoff-Chu}\\{\fontfamily{ptm}\selectfont modulator}};
	\draw (229,62) node [anchor=north west][inner sep=0.75pt]   [align=left] {$\displaystyle r$};
	\draw (267,62) node [anchor=north west][inner sep=0.75pt]   [align=left] {$\displaystyle N$};
	\draw (303,103) node [anchor=north west][inner sep=0.75pt]   [align=left] {$\displaystyle x_{d}( \ell T+kT_{c})$};
	\draw (412,117) node [anchor=north west][inner sep=0.75pt]   [align=center] {$\displaystyle g( \cdot )$};
	\draw (381,154) node [anchor=north west][inner sep=0.75pt]   [align=left] {\begin{minipage}[lt]{61.08pt}\setlength\topsep{0pt}
			\begin{center}
				{\fontfamily{ptm}\selectfont pulse shaping }\\{\fontfamily{ptm}\selectfont filter}
			\end{center}
			
	\end{minipage}};
	\draw (466,80) node [anchor=north west][inner sep=0.75pt]   [align=left] {{\fontfamily{ptm}\selectfont baseband}\\{\fontfamily{ptm}\selectfont equivalent }\\{\fontfamily{ptm}\selectfont signal}};
	\draw (558,106) node [anchor=north west][inner sep=0.75pt]   [align=left] {{\fontfamily{ptm}\selectfont up conversion}\\{\fontfamily{ptm}\selectfont RF front-end}};

\end{tikzpicture}
\caption{The Golden Modulator}
\label{fig:single-link-tx}
\end{figure*}

To be transmitted, the discrete-time GM signals must be converted in the continuous time domain through a pulse shaping filter $g(t)$. The complete chain to produce the analog continuous time GM modulated signal is shown in Fig.~\ref{fig:single-link-tx}.

\begin{figure}[tbh]
	\centering
	\includegraphics[width=8cm, height=8cm]{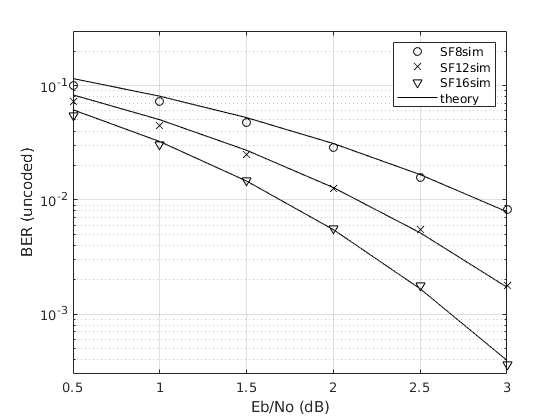}
	\caption{Theoretical and simulated BER for the Golden Modulation for \linebreak $SF=8,12,16$. }
	\label{fig:berebnogmsf8sf12sf16}
\end{figure}

A practical choice for the pulse shaping filter $g(t)$ typically results in a signal bandwidth $B_{ch}>\frac{1}{T_c}$. For example, for the widely used root raised cosine waveform, we have $B_{ch}= (1+\beta)B$, where $\beta$ is the roll-off factor and determines the excess bandwidth.

The performance of the Golden modulation in an AWGN channel is provided in Fig. \ref{fig:berebnogmsf8sf12sf16} for the case of ideal Nyquist filtering, where  the uncoded Bit Error Rate (BER) obtained via computer simulations is compared to that obtained via the analytical approximation of Equation (\ref{eq:ber_lora}). Simulation results performed for Golden Modulation with SF 8, 12 and 16 show a very good match with the analytical  curves.

\begin{figure}[t]
	\centering
	\includegraphics[width=1.1\linewidth]{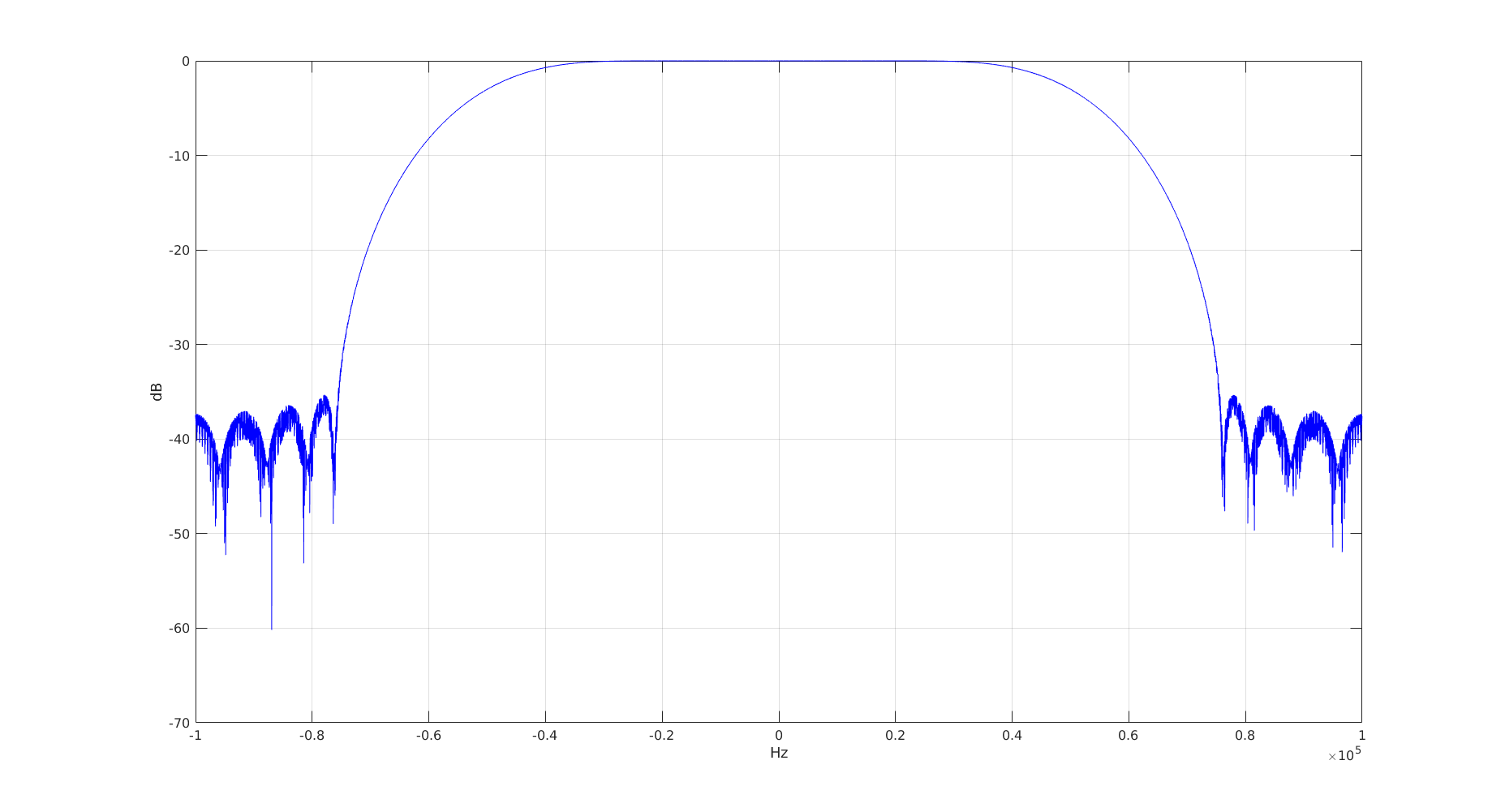}
	\caption{Power spectral density of the Golden Modulation for a bandwidth of 100 kHz and SF=12 }
	\label{fig:gmspectrum}
\end{figure}

To conclude the characterization of the Golden Modulation, in Fig. \ref{fig:gmspectrum} the simulation result of the Power Spectral Density for a bandwidth of  100 kHz and SF12 is shown.

\color{black}

\section{The receiver}
\label{sec:receiver}

Ignoring for the sake of simplicity the effect of the transmission channel and assuming $N$ odd, to demodulate the  symbol transmitted in the period $\ell T$ we consider the received signal 
\begin{equation}
	y(\ell T + kT_c) \stackrel{\text{def}}{=}  e^{\jmath \frac{\pi}{N} r \left(k+1+2q_\ell\right)k} +w(\ell T + kT_c) 
	\label{ref:rec_sig}
\end{equation}
where $w(\cdot)$ is a Complex Additive White Gaussian Noise ($\mathcal{C}$AWGN), with variance $\sigma_w^2$. Since the receiver is assumed to know the parameters $r$ and $N$, it can compute the signal

\begin{equation}
\begin{split}
	y_{m}(\ell T + kT_c) & \stackrel{\text{def}}{=}  \left[e^{\jmath \frac{\pi}{N} r \left(k+1+2q_\ell\right)k} +w(\ell T + kT_c)\right] \cdot \\
	& \quad \cdot e^{-\jmath \frac{\pi}{N} r \left(k+1\right)k}\\
	& = e^{\jmath \frac{2\pi}{N} q_\ell k} +w_m(\ell T + kT_c)
\end{split}
\end{equation}
where $w_m(\ell T + kT_c) \stackrel{\text{def}}{=} w (\ell T + kT_c)  \cdot e^{-\jmath \frac{\pi}{N} r \left(k+1\right)k} $ is also $\mathcal{C}$AWGN with variance $\sigma_w^2$.

The value of $q_\ell$ can be obtained by forming the $N$ dimensional vector
\begin{equation}
	\mathbf{Y}(\ell T) = \begin{bmatrix}
		y_{m}(\ell T + 0 \cdot T_c) \\
		y_{m}(\ell T + 1  \cdot T_c) \\
		\vdots \\
		y_{m}(\ell T + (N-1) \cdot T_c) 
	\end{bmatrix},
\end{equation}
computing $\mathbf{Z}(\ell T) = \text{DFT}(\mathbf{Y})$, and taking the index of the entry with largest magnitude. Notably, if $N$ is prime the DFT can still be made at a complexity similar to the usual radix--2 one via the FFTW algorithm \cite{FFTW}. This complexity is the same as in LoRa systems.

\section{Golden Modulation: multiple access and multiple rate}
\label{sec:multiple-links}
From Section \ref{sec:single-link}, one can see that for any link employing the Golden Modulation we have three degrees of freedom (by considering fixed both the  bandwidth $B$ and the pulse shaping filter), namely: \begin{itemize}
	\item the root $r$;
	\item the length $N$; and
	\item the cardinality of the modulation $M=2^b < N$.
\end{itemize}

We suppose, as a first step, to fix the length $N$ as odd and the cardinality $M=2^b < N$. This is equivalent to fixing the rate of the information source whose output is carried in the link. 
We can see from Eq. (\ref{eq:2}) that for the transmission of the symbols we still have one degree of freedom, i.e., the root $r$, which can be used to separate transmissions issued \color{black} by different end devices, as discussed in the following.  

\subsection{Multiple users with the same Spreading Factor }
\label{subsec:multiple-sync-sources}
In this subsection we show how we can multiplex in a single channel of band $B_{ch}$ multiple users with the same rate.
 
We know from Section \ref{sec:sequences} that, by selecting $N$ odd and any pair of distinct roots $r_1, r_2$ such that $\gcd(r_1-r_2,N)=1$, the  signals  in the time interval $\left\{\ell T, (\ell+1) T_c, \ldots + (\ell+N-1)T_c)\right\}$, namely
$x(\ell T + kT_c;q_\ell^{(1)},N;r_1)$ and $x(\ell T + kT_c;q_\ell^{(2)};N;r_2)$, have their cyclic cross-correlation constant and equal to $\frac{1}{\sqrt{N}}$. 

Given $N$ prime, a set of $K\leq N-1$ distinct roots $\mathcal{R}_{N,K} = \left\{r_1, r_2, \ldots r_K\right\}$ can be selected such that   $\gcd(r_i-r_j,N)=1$ for $r_i,r_j \in \mathcal{R}_{N,K}, i \neq j$. Any link such as the one described in Section \ref{sec:single-link}, using the root $r_i \in  \mathcal{R}_{N,K}$,  is then ``almost'' orthogonal to  any other link of the same type using a different root $r_j \in  \mathcal{R}_{N,K}$, $i \neq j$. More precisely, the signals $x(nT_c;q_\ell^{(i)}, N;r_i) $  and $x(nT_c;q_\ell^{(j)},N;r_j)$ have a low (equal to $\frac{1}{\sqrt{N}}))$ cyclic cross-correlation and the symbols $q_\ell^{(i)}$ and $q_\ell^{(j)}$ transmitted by the two links can be recovered by a suitable receiver (described in Section \ref{sec:receiver}) for SNR in the region of operation of the system. This property makes it possible to define up to $N-1$ different links, thereby enabling massive access in IoT systems.

In a given frequency channel, a user is characterized by its root  $ r $ and  its spreading factor $ SF $, and the level of interference between users directly depends on the circular cross-correlation between their respective signals  $x(nT_c;q_\ell^{(i)}, N;r_i) $  and $x(nT_c;q_\ell^{(j)}, N;r_j)$. Golden Modulation is designed to exploit the conditions for perfect auto-correlation and minimal cross-correlation properties of Zadoff-Chu sequences  equal respectively to  $1$ and $1/\sqrt{N}$.   

\begin{figure}[t]
	\centering
	\includegraphics[width=8cm, height=8cm]{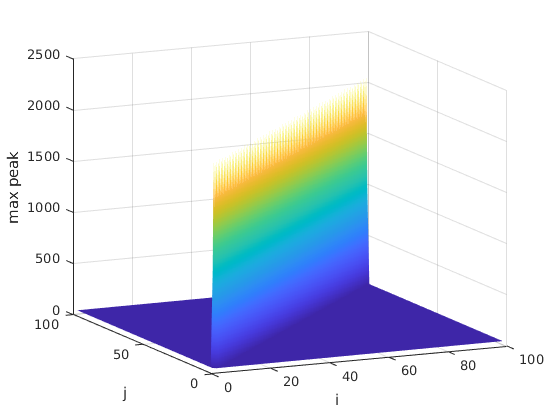}
	\caption{Cyclic cross-correlation maximum peaks for $N=2053$ Zadoff-Chu sequences  $Z_N^{r_i,0}(k)$ and $Z_N^{r_j,0}(k)$ (subset of 100 roots); indices $i$ and $j$ are plotted on the two horizontal axes.}
	\label{fig:zccrosscorr}
\end{figure}

The above performance applies identically to the full set of Zadoff-Chu sequences of length $N$ as illustrated by Fig.~\ref{fig:zccrosscorr}, which shows the cross-correlation results between any two sequences from  a set of 100 Zadoff-Chu sequences of length 2053. The set of "quasi orthogonal" Golden Modulation signals to be used by devices is hence very large (up to $N-1$) and increases with the spreading factor. Similarly, the orthogonality level between users increases with the spreading factor as shown in Table \ref{tab:table1}, where ``interference rejection'' is the highest value of the cross-correlation between a symbol with a certain SF and any other symbol with the same SF.\footnote{For each Spreading Factor $SF$, $N$ was chosen as the smallest prime number greater than or equal to $2^{SF}$. It is also possible to truncate such sequence to a length exactly equal to $2^{SF}$, for compatibility with existing systems, as discussed in Section \ref{sec:higher_layers_networks}.}

\begin{table}[t]
  \begin{center}
    \caption{Number of Golden Modulation waveforms and interference levels per SF.}
    \label{tab:table1}
    \begin{tabular}{c|c|c} 
      \textbf{Spreading} & \textbf{Interference Rejection} & \textbf{Waveforms Set Size }\\
      $SF$ & $dB$ & $N-1$ \\
      \hline
      7 & -10.6 & 130\\
      8 & -12.0 & 256\\
      9 & -13.6 & 520\\
      10 & -15.1 & 1030\\
      11 & -16.6 & 2052\\
      12 & -18.1 & 4098\\
      13 & -19.6 & 8208\\
      14 & -21.1 & 16410\\
      15 & -22.6 & 32770\\
      16 & -24.1 & 65536\\

    \end{tabular}
  \end{center}
\end{table}

A remark on Figure~\ref{fig:zccrosscorr} is in order. As expected, when the two sequences are the same, i.e., $i=j$, we observe a large peak equal to $N$. In all other cases (i.e., $i\neq j$) the maximum peak is much smaller (with a value of $\sqrt{N}$) and independent of $i$ and $j$ (i.e., it is flat, as can be observed from Figure~\ref{fig:zccrosscorr}).

\begin{figure}[t]
	\centering
	\includegraphics[width=0.9\linewidth,height=10cm]{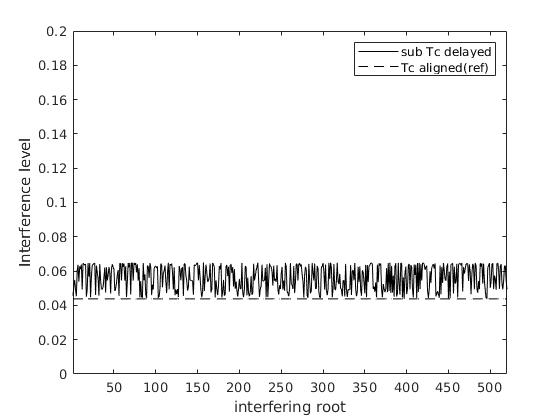}
	\caption{Crosscorrelation maxima between symbols of an $ SF=9 $ link  and any uncoordinated (not time aligned) interfering link with  $ SF=9 $ and a different root  }
	\label{fig:gmcorroutsf9uncoord}
\end{figure}

It must be noted that comparable performance can be obtained  when considering two ``uncoordinated'' links, i.e., links that are not aligned in time. In this case, since signals issued by different users reach the receiver with independent timing, the underlying Zadoff-Chu sequences of the different links  overlap asynchronously and interference will appear  at the correlator output, depending on their aperiodic cross-correlation. Correlation properties of polyphase sequences have been extensively studied in the literature, e.g., see \cite{Welch1974}, \cite{Antweiler1990}, \cite{Mow1992} and \cite{Fan1994}, and \cite{Mow1997} eventually derived the aperiodic cross-correlation bounds for Zadoff-Chu sequences. These bounds show the very good aperiodic correlation properties of Zadoff-Chu sequences, as illustrated in the example of Fig. \ref{fig:gmcorroutsf9uncoord} showing the maximum interference level experienced by a target link with $SF=9$ due to all possible other uncoordinated interfering signals with $SF=9$. 

\begin{figure}[t]
	\centering
	\includegraphics[width=\linewidth]{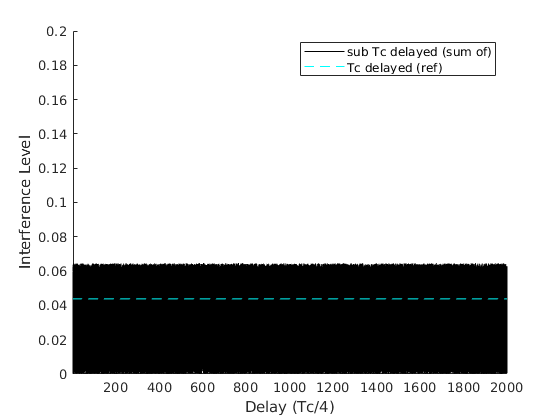}
	\caption{Crosscorrelation between symbols of uncoordinated $ SF=9 $ links depending on delays (100 randomly selected interferers with random sub Tc delays)}
	\label{fig:gmcorroutsf9uncoorddelays}
\end{figure}

Fig. \ref{fig:gmcorroutsf9uncoorddelays} shows that the maximum interference level between different GM signals with the same SF is almost independent of the fractional $T_c$ delay.
In particular, to study the effect of fractional misalignment between the two sequences, we plotted the cross-correlation values for all time-shifts that are multiples of $T_c/4$. While for time-shifts multiples of $T_c$ the normalized interference level is $1/\sqrt{N}$ ($\simeq 0.438$ for $N=521$ in this case, represented by the horizontal dashed line), for fractional time-shifts the possible values of the interference level vary within the range from 0 to about 0.65, showing that fractional misalignment has a modest price in terms of a (uniformly bounded) interference increase. 

Golden Modulation leverages these low cross-correlation properties of the filtered Zadoff-Chu sequences to provide ultra-high capacity without compromising range or power consumption. 
This is a major advantage from a system point of view since multiple devices can transmit simultaneously using different Zadoff-Chu roots without the need for any synchronization between the different transmissions, hence avoiding the need for complex and energy consuming synchronization mechanisms. This permits to increase by several orders of magnitude the system capacity of IoT networks without paying any penalty in terms of single link performance, power consumption, and receiver complexity.

\begin{figure}[tbh]
	\centering
	\includegraphics[width=\linewidth]{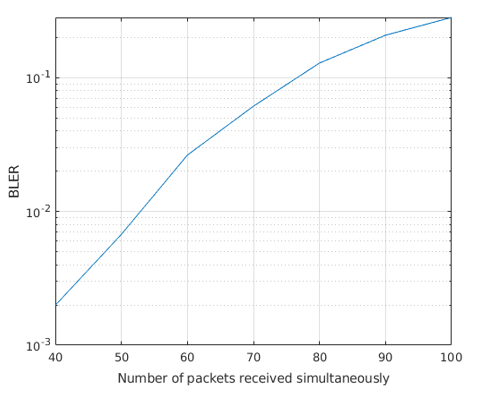}
	\caption{Impact of Multiuser Interference on GM demodulation for $ SF=12 $ }
	\label{fig:gmmultiuserbler}
\end{figure}

The performance of GM in a multi user configuration is presented for SF 12 in Fig. \ref{fig:gmmultiuserbler}, where packets from the different users are received with equal power and random delays. 
It can be seen that the demodulation performance of Golden Modulation remains very good in the presence of several tens of simultaneous interfering devices while LoRa is limited to the reception of a single packet per SF (provided that it is free of collisions). 

Last but not least, this major capacity boost is obtained without the need for advanced receivers thanks to the ``quasi orthogonality" of the frames received from the different users. In particular, the received signals can be demodulated by single user receivers without the need for complex interference cancellation techniques.
More precisely, we can state that, considering two links with the same rate  $x(nT_c;q_{\ell(n)}^{(i)},N;r_i)$ and $x((n+n_0)T_c;q_{\ell(n)}^{(j)},N;r_j)$ with $0 \le n_0 < N$ and $r_i\neq r_j$, the respective sequences of symbols $q_{\ell(n)}^{(i)}$ and $q_{\ell(n)}^{(j)}$ transmitted by the two links can be recovered by a suitable receiver without the need for costly interference cancellation algorithms nor complex channel coding schemes.

\subsection{Multiple  users with different Spreading Factors}
\label{subsec:multiple-async-sources}
In this subsection we consider multiple users with different Spreading Factors, to explain the case when two  users may use two generic lengths $N_1$ and $N_2$, not necessarily equal,  using roots $r_i \in \mathcal{R}_{N_1,K_1}$ and $r_j \in \mathcal{R}_{N_2,K_2}$. The generic two users may also be uncoordinated, i.e., not aligned in time. 
We note that this case corresponds to the actual situation where End-Nodes (ED) in a Low Power Wide Area Network (LPWAN) autonomously trigger their transmissions at the rate suitable for their respective link budget and the time suitable for their respective application.

\begin{figure}[tbh]
	\centering
	\includegraphics[width=8cm, height=8cm]{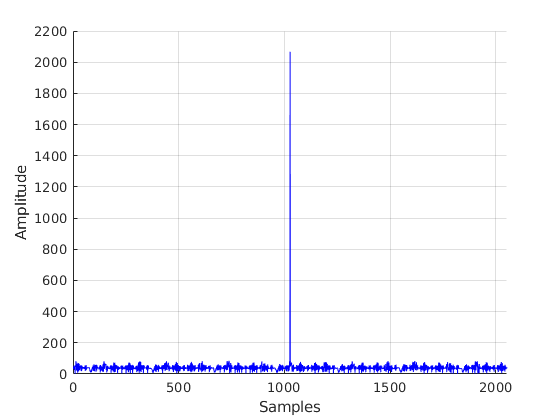}
	\caption{Correlator output for an $ SF = 11 $ link in the presence of an uncoordinated interfering link with $ SF = 10 $}
	\label{fig:gmcorroutdiffsf2}
\end{figure}

Golden Modulation, relying on the properties of its underlying Zadoff-Chu sequences, exhibits a ``quasi-orthogonality" between signals with different Spreading Factors as well, whichever their roots. As an example, the interference generated by a signal with different Spreading Factor (i.e., different $N$) is shown in Fig. \ref{fig:gmcorroutdiffsf2}, which plots the output of the receiver that results from the superposition of an intended signal and an uncoordinated interferer operating with $ SF=11$ and $SF=10 $, respectively. The intended signal produces the expected peak of value $N=2053$ at the output of the correlator according to its time-shift (around 1000 samples in this case) and zero autocorrelation otherwise, whereas the interferer produces low cross-correlation values for all possible time-shifts, confirming the good cross-correlation properties of these waveforms even for different spreading factors.

\section{Golden Modulation Networks: Higher-Layer Protocols and Networking }
\label{sec:higher_layers_networks}

\begin{figure*}[b]
	\centering
 \begin{tikzpicture}
	\draw[gray, thick,<-> ] (-0.75,2) -- (14.25,2);
	\node at (7,2.2) {FRAME};
	\draw[gray, thick,<-> ] (-0.75,1) -- (5.25,1);
	\node at (2.5,1.2) {PREAMBLE};
\draw[gray, thick,<-> ] (5.25,1) -- (14.25,1);
\node at (9.5,1.2) {PAYLOAD};
	\node[rectangle, 
	draw = blue,
	text = blue,
	fill = gray!10,
	minimum width = 1.5cm, 
	minimum height = 1cm] (r1) at (0,0) {$Z_N^{r,q=q_0}$};
	\node[rectangle, right= 0.75cm of r1.center,
	draw = blue,
	text = blue,
	fill = gray!10,
	minimum width = 1.5cm, 
	minimum height = 1cm] (r2) {$Z_N^{r,q=q_0}$};
	\node[rectangle, right= 0.75cm of r2.center,
	draw = blue,
	text = blue,
	fill = gray!10,
	minimum width = 1.5cm, 
	minimum height = 1cm] (r3) {$\cdots$};
	\node[rectangle, right= 0.75cm of r3.center,
	draw = blue,
	text = blue,
	fill = gray!10,
	minimum width = 1.5cm, 
	minimum height = 1cm] (r4) {$Z_N^{r,q=q_0}$};
	\node[rectangle, right= 0.75cm of r4.center,
	draw = black,
	text = black,
	minimum width = 1.5cm, 
	minimum height = 1cm] (r5) {$Z_N^{r,q(1 T)}$};
	\node[rectangle, right= 0.75cm of r5.center,
	draw = black,
	text = black,
	minimum width = 1.5cm, 
	minimum height = 1cm] (r6) {$Z_N^{r,q( 2T)}$};
	\node[rectangle, right= 0.75cm of r6.center,
	draw = black,
	text = black,
	minimum width = 1.5cm, 
	minimum height = 1cm] (r7) {$Z_N^{r,q( 3T)}$};
	\node[rectangle, right= 0.75cm of r7.center,
	draw = black,
	text = black,
	minimum width = 1.5cm, 
	minimum height = 1cm] (r8) {$Z_N^{r,q( 4T)}$};
	\node[rectangle, right= 0.75cm of r8.center,
	draw = black,
	text = black,
	minimum width = 1.5cm, 
	minimum height = 1cm] (r9) {$\cdots$};
	\node[rectangle, right= 0.75cm of r9.center,
	draw = black,
	text = black,
	minimum width = 1.5cm, 
	minimum height = 1cm] (r10) {$Z_N^{r,q( LT)}$};
	
\end{tikzpicture}
	\caption{Example of Golden Modulation frame format}
	\label{fig:gmframe}
\end{figure*}
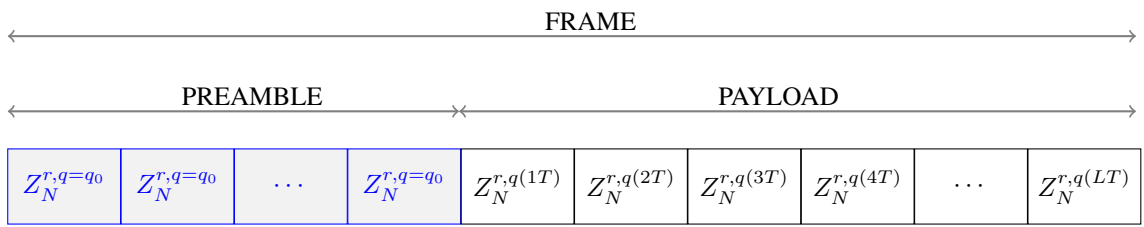

The Golden Modulation frame is made of a series of concatenated Zadoff-Chu sequences covering a preamble and a payload section (see the Golden Modulation frame format in Fig. \ref{fig:gmframe}). Frames issued by different devices can advantageously use different roots or sets of roots; this allows easy multiuser reception without the need for device coordination. It must be highlighted that there are $ N-1 $ possible roots for Zadoff-Chu sequences of length $ N $, hence providing the potential for the generation of thousands of ``quasi orthogonal" frames with the same or different $ SF $,  totally unlocking the IoT networks capacity. In addition, the ability to support the highly desirable uncoordinated mode of operation guarantees robustness and extremely low protocol overhead.     

\begin{figure}[t]
	\centering
	\includegraphics[width=8cm, height=8cm]{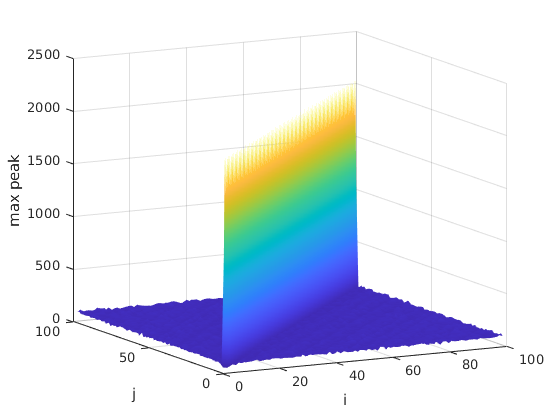}
    \caption{Cyclic cross-correlation maximum peaks for the $ N=2053 $ Zadoff-Chu sequences $Z_N^{r_i,0}(k)$ and $Z_N^{r_j,0}(k)$ truncated to $ N=2048 $ (subset of 100 roots); indices $i$ and $j$ are plotted on the two horizontal axes. }
	\label{fig:zctruncxcorr}
\end{figure}

It must be noticed that, despite its baseline design using odd-length Zadoff-Chu sequences, Golden Modulation can also be implemented based on even-length sequences. For example, an implementation based on truncated odd length Zadoff-Chu sequences leading to an effective sequence length equal to a power of 2 is also possible. In the latter case, the degradation of GM's orthogonality is marginal, as shown in Fig. \ref{fig:zctruncxcorr}, and has no noticeable effect on its performance. Again, looking at Figure~\ref{fig:zctruncxcorr}, we observe that for $i\neq j$ the maximum peak is small and almost independent of the indices of the two sequences.

An implementation based on truncated odd-length Zadoff-Chu sequences leading to an effective sequence length equal to a power of 2 is interesting for cases where adaptation to processing units or to size-constrained lower layer protocols  is required. The latter case is actually very much of practical use. For example, it is straightforward to implement the LoRaWAN protocols on top of  GM  with Zadoff-Chu truncated sequences to achieve an effective sequence length equal to a power of 2. As a matter of fact, once the value of the SF for the LoRaWAN is selected, the same SF can be selected for GM. Moreover, once a certain bandwidth $B$ is selected for the LoRaWAN protocol, the same bandwidth $B$ can be selected for GM. All of the specifications (for example~\cite{lorawan_l2}) refer to bytes for the fields of any message and the messages can then be immediately translated in bytes carried by GM. As one can notice from Figure~\ref{fig:gmframe}, the structure (preamble followed by data) is the same. All the "regional parameters" as per~\cite{lorawan_rp} are automatically reusable (for the same bandwidth $B$), ensuring compliance with the regulations worldwide. In general, we can safely state that the complete LoRaWAN system of specifications (including the backend, the Firmware Update Over The Air (FUOTA), etc.) becomes completely re-usable. For the commercial viability of GM, this represents a decisive strength. Of course, here LoRaWAN is just an example, although important, and the byte oriented transmission made possible by the use of an effective sequence length equal to a power of 2 opens the possibility to use GM in many other systems.

\section{Conclusion}
\label{sec:conclusions}

In this paper, we introduced Golden Modulation, a novel modulation with ultra-long range capability and massive capacity. The basic modulation principles relying on spectrum spreading via direct Zadoff-Chu sequences modulation have been presented and the corresponding theoretical BER performance in AWGN channel has been derived and compared by simulation with realistic Golden Modulation receiver performance. A basic low complexity DFT-based receiver has been described. The fundamentals of multi-user operation have been described and the corresponding multi-user interference evaluated from both a theoretical and a simulation point view. The very low multi-user interference level, even in uncoordinated conditions, coupled with the numerous available waveforms, unlocks the well-known capacity bottleneck of IoT networks. Overall, low complexity receiver, flexible frame structure and minimal protocol constraints pave the way for the deployment of LPWAN networks with truly massive capacity, or for the upgrade of existing ones. Furthermore, the possibility to use an effective sequence length equal to a power of 2 makes GM suitable as a physical layer for many existing protocol suites. 

{\appendix[Closed form expression of \\ Golden Modulation BER]} 
\label{subsec:gm-ber}
According to Equation (\ref{eq:orthogonality}), the discrete-time signal space for the Golden Modulation admits as an orthonormal basis the set of sequences $\frac{1}{\sqrt{N}}x(kT_c;q,N;r)$ indexed by the possible transmitted symbol $q \in \{0,1, \ldots, M-1\}$ and for the running time index $kT_c$, $k \in \{0,1, \ldots, N-1\}$. One can easily see that the detection problem is the same (apart from the notation) as in \cite{Robert2018}
As demonstrated in \cite{Robert2018}, then the bit error probability can be expressed as   

\begin{equation}
\ P_{b}=\int_{0}^{\infty} {\left[1-\left[1-\exp\left[-\frac{\beta^{2}} {2\,\sigma^{2}}\right]\right]^{2^{SF-1}}\right]f_{\beta}(\beta) \, d\beta }
\end{equation}
where $\sigma^{2} = \frac {N_{0}} {2}$, with $N_{0}$ the single-sided noise power spectral density, and $f_{\beta}(\beta)$ is the probability density function for the Rician distributed $\beta$ with shape parameter $\kappa_{\beta} = \frac {E_{S}} {N_{0}} = \frac{SF}{N_0}$.
Further noticing the very high spreading gains used and applying Gaussian approximations,\cite{Robert2018} derives an approximation of the corresponding bit error probability valid for high spreading gains     
\begin{equation} 
\ P_{b} \approx 0.5 Q(\sqrt{\Gamma.2^{SF+1}}-\sqrt{1.386.SF +1.154}) \ 
\label{eq:robert_approx}
\end{equation}
where $\Gamma$ is the SNR. This expression is also valid for GM.
Corroborating this statement, the simulations performed with $ SF=8,12,16 $ show a very good match with the curves predicted by Equation~(\ref{eq:robert_approx}), as shown in Fig.~\ref{fig:berebnogmsf8sf12sf16}.

\section*{Acknowledgments}
The authors would like to thank Prof. Fabien Ferrero from University of Nice (UCA-LEAT) for the constructive discussions and for his support for the experimental validations of the Golden Modulation performed in his lab premises. This work was partially supported by the European Union under the Italian National Recovery and Resilience Plan (NRRP) of NextGenerationEU, partnership on ``Telecommunications of the Future” (PE0000001 - program
``RESTART”).

\bibliographystyle{IEEEtran}
\bibliography{golden}



\balance

\vspace{-33pt}
\begin{IEEEbiographynophoto}{Lorenzo Vangelista}
(S’93-M’97-SM’02) received the Laurea and Ph.D. degrees in electrical and telecommunication engineering from the University of Padova, Padova, Italy, in 1992 and 1995, respectively.He subsequently joined the Transmission and Optical Technology Department, CSELT, Torino,Italy. From December 1996 to January 2002, he was with Telit Mobile Terminals, Trieste, Italy, and then, until May 2003, he was with Microcell A/S, Copenaghen, Denmark. In July 2006, he joined the Worldwide Organization of Infineon Technologies as Program Manager. From October 2006 to October 2021 he has been an Associate Professor of Telecommunication with the Department of Information Engineering, Padova University, Padova, Italy, where he is now Full Professor. His research interests include signal theory, multicarrier modulation techniques, cellular networks and Internet of Things connectivity with special focus on Low Power Wide Area Networks.
\end{IEEEbiographynophoto}
\vspace{-33pt}
\begin{IEEEbiographynophoto}{Bruno Jechoux}
received the Electrical Engineering M.Sc. from Centrale-Supelec engineering school in 1994 with a specialty in Radiocommunications. He has more than 20 years of experience in the design and standardization of advanced wireless systems including WiFi, LTE and 5G. Prior to co-founding Ternwaves he has been with Mitsubishi, Motorola, Infineon, Intel and TCL. From 2011 to 2016  he was with Intel where he was principal wireless system engineer in charge of defining the 4G and 4.5G modems architecture. From 2016 to 2021 he was head of TCL 5G European research Lab (Sophia Antipolis, France) in charge of 5G  standardization and advanced 5G demonstrator (world’s first 5G full software implementation at Mobile World Congress 2018). He is the author of more than 50 patents.
\end{IEEEbiographynophoto}
\vspace{-33pt}
\begin{IEEEbiographynophoto}{Jean-Xavier Canonici}
received the Electrical Engineering M.Sc. from Centrale-Supelec engineering school in 1994 with a with a specialization in Digital communications and Electronics.  He has 25 years of experience in modem definition and development for fixed wireless, satellite, cellular 3G/4G \& 5G-IOT. His experience was built in the following  companies: Intel Mobile Communications, Sequans, Infineon, Siemens, Alcatel Telspace, Alcatel CIT and CEA. From 2011 to 2016, he was within Intel as Principal Engineer for Wireless Systems and Head of Modem System Engineering team working more specifically on high-end smartphone platforms for lead customers like Apple or Samsung. He was just before the creation of Ternwaves Principal Physical Layer Algorithms \& Technical Manager at Sequans in charge of NB-IoT modem for CatM1/CatNB1 Platforms. He holds more than 10 US patents.
\end{IEEEbiographynophoto}
\vspace{-33pt}
\begin{IEEEbiographynophoto}{Michele Zorzi}
(S’89-M’92-SM’97-F'07) received the Laurea and Ph.D. degrees in electrical engineering from the University of Padova, Italy, in 1990 and 1994, respectively. From 1992 to 1993, he was on leave at the University of California at San Diego (UCSD). In 1993, he joined the Faculty of the Dipartimento di Elettronica e Informazione, Politecnico di Milano, Italy. After spending three years with the Center for Wireless Communications, UCSD, he joined the School of Engineering, University of Ferrara, Italy, in 1998, where he became a Professor in 2000. Since November 2003, he has been on the Faculty of the Information Engineering Department, University of Padova. His current research interests include performance evaluation in mobile communications systems, the Internet of Things, cognitive communications and networking, 5G mmWave cellular systems, vehicular networks, and underwater communications and networks. He has served/will serve the IEEE Communications Society as a Member-at-Large of the Board of Governors from 2009 to 2011 and from 2021 to 2023, as the Director of Education from 2014 to 2015, and as the Director of Journals from 2020 to 2021. He received several awards from the IEEE Communications Society, including the Best Tutorial Paper Award in 2008 and 2019, the Education Award in 2016, the Stephen O. Rice Best Paper Award in 2018, and the Joseph LoCicero Award for Exemplary Service to Publications in 2020. He was the Editor-in-Chief of the IEEE Wireless Communications Magazine from 2003 to 2005, the IEEE Transactions on Communications from 2008 to 2011, and the IEEE Transactions on Cognitive Communications and Networking from 2014 to 2018.
\end{IEEEbiographynophoto}

\vfill

\end{document}